\documentclass[english,aps,prb,showpacs,twocolumn]{revtex4-2}
\usepackage[T1]{fontenc}
\usepackage[latin9]{inputenc}
\usepackage{babel}
\usepackage{graphicx}
\usepackage{bm}
\usepackage{amsmath}
\usepackage{amssymb}
\usepackage{amsfonts}
\usepackage{dcolumn}
\usepackage{bbold}
\usepackage{tikz}
\usepackage{xfp}

\usepackage[margin=15mm]{geometry}
\colorlet{linkequation}{blue}
\usepackage[colorlinks=true,linkcolor=blue,citecolor=blue,urlcolor=blue]{hyperref}
\usetikzlibrary{shadings}

\usepackage{color}
\newcommand{\blue}{\color{blue}}

\begin{document}

%----------------------------------------------------------------------

\date{\today}
\begin{abstract}

Spin-nutation resonance has been well-explored in one-sublattice ferromagnets. Here, we investigate the spin nutation in two-sublattice antiferromagnets {as well as,} {for comparison,} ferrimagnets with 
inter- {and intra-}sublattice nutation coupling. In particular, we derive the % 
{susceptibility} of the two-sublattice magnetic system {in response} to an applied external {magnetic} field. % 
To this end, the antiferromagnetic and ferrimagnetic {(sub-THz)} precession and {THz} nutation resonance frequencies are calculated. Our results show that the precession resonance frequencies and effective damping decrease with \textit{intra}-sublattice nutation {coupling}, while {they} increase with \textit{inter}-sublattice nutation in an antiferromagnet. {However, we find that the {THz} nutation resonance frequencies decrease with {both the {\it intra} and {\it inter}}-sublattice nutation couplings. {For ferrimagnets, conversely, we calculate two nutation modes with distinct {frequencies}, unlike antiferromagnets.}  The exchange-like precession resonance frequency of ferrimagnets decreases with {\it intra}-sublattice nutation coupling and increases with {\it inter}-sublattice nutation coupling, like antiferromagnets, but the ferromagnetic-like precession frequency of ferrimagnets is practically invariant to the} 
{\it intra} and {\it inter}-sublattice nutation couplings.

\end{abstract}
%----------------------------------------------------------------------

\title{{Influence of} inter-sublattice {coupling on the {terahertz} nutation spin dynamics} in antiferromagnets}

\author{Ritwik Mondal$^{1,2}$}
\email[]{mondal@fzu.cz}
\author{Peter M. Oppeneer$^1$}
\affiliation{$^1$Department of Physics and Astronomy, Uppsala University, Box 516, Uppsala, SE-75120, Sweden}
\affiliation{ $^2$Department of Spintronics and Nanoelectronics, Institute of Physics of the Czech Academy of Sciences, Cukrovarnick\'a 10, CZ - 162 00 Praha 6, Czech Republic}

\maketitle

\section{Introduction}
{E}fficient spin manipulation at ultrashort timescales holds promise for {applications in} future magnetic memory technology~\cite{Bigot1996,Stanciu2007,Vahaplar2009,CARVAbook,John2017}. 
Introduced by Landau and Lifshitz, the time evolution of magnetization $\bm{M}(\bm{r},t)$, {can be} described by {the} phenomenological Landau-Lifshitz-Gilbert (LLG) equation of motion, {which} reads~\cite{landau35,Gilbert1955,gilbert56,gilbert04}
\begin{align}
\dot{\bm{M}} & = -\gamma \left(\bm{M}\times \bm{H}\right) + \frac{\alpha}{M_{0}}\left(\bm{M}\times \dot{ \bm{M}}\right) ,
\end{align}
with the gyromagnetic ratio $\gamma$,  constant magnetization amplitude $M_0$, and Gilbert damping parameter $\alpha$.
The LLG equation consists of the precession of spins around a field $\bm{H}$ and transverse damping that aligns the spins towards the field direction. While the spin precessional motion can be explained by Zeeman-like field-spin coupling, there are several fundamental and microscopic mechanisms leading to Gilbert damping~\cite{kambersky70,kambersky76,Koreman1972PRB,kunes02,kambersky07,tserkovnyak02,hickey09,gilmore07,Mondal2016,Mondal2018PRB,Mondal2018JPCM,Mondal2015a,Mondal2015b}.

{When one approaches the} femtosecond regime, {however}, the spin dynamics can {\it not only} be described by the traditional {LLG} dynamical equation of motion \cite{Mondal2019PRB,mondal2021terahertz}, but {it} has to be supplemented by a fast dynamics {term} due to magnetic inertia
\cite{Ciornei2011,Ciornei2010thesis,Wegrowe2012}.    
Essentially, the inclusion of magnetic inertia leads to a spin nutation at  ultrashort timescales and can be described by a torque due to {a} double time-derivative of the magnetization i.e., $\bm{M}\times \ddot{\bm{M}}$~\cite{Ciornei2010thesis,Bottcher2012}. The inertial LLG (ILLG) equation of motion has the form
\begin{align}
\dot{\bm{M}} & = -\gamma \left(\bm{M}\times \bm{H}\right) + \frac{\alpha}{M_{0}}\left(\bm{M}\times \dot{ \bm{M}}\right) + \frac{\eta}{M_{0}}\left(\bm{M}\times \ddot{ \bm{M}}\right) ,
\end{align}   
with the inertial relaxation time $\eta$. In general, the Gilbert damping $\alpha$ and the inertial relaxation time $\eta$ are tensors~\cite{Mondal2017Nutation}, however, for an isotropic system, these parameters can be considered as scalars. The emergence of spin nutation has been attributed to {an} extension of Kambersk\'y breathing Fermi surface model~\cite{Fahnle2011JPCM,Fahnle2011}, {namely, an} $s-d$-like interaction spin model between local magnetization and itinerant electrons~\cite{Bhattacharjee2012,Utkarsh2019}. Moreover, the ILLG equation has been derived from {the} fundamental Dirac equation~\cite{Mondal2017Nutation,Mondal2018JPCM}. Note that the Gilbert damping and inertial relaxation time are related to each other as the Gilbert damping is associated with the imaginary part of the susceptibility, while the inertial dynamics are associated with the real part of the susceptibility~\cite{Mondal2018JPCM,Thonig2017}. The characteristic timescales of the nutation have been predicted to {be in} a range of $1-100$ fs~\cite{Ciornei2011,Bhattacharjee2012,Li2015,Makhfudz2020} and $1 -10$ ps~\cite{Makhfudz2020,neeraj2019experimental}. More recently, it has been demonstrated that simple classical mechanical considerations superimposed with Gilbert dynamics naturally lead to magnetic inertial dynamics \cite{Giordano2020,Titov2021}.

Theoretically, the spin nutation has {recently} been extensively discussed {for} one-sublattice ferromagnets~\cite{Ciornei2011,Makhfudz2020,cherkasskii2020nutation,Cherkasski2021,lomonosov2021anatomy,Rahman_2021,TitovPRB2021}. The nutation resonance has also been observed in experiments, however {for} two-sublattice ferromagnets~\cite{neeraj2019experimental}. {A recent} theoretical investigation predicts that the precession and nutation resonance frequencies may overlap in two-sublattice ferromagnets~\cite{MondalJPCM2021}.  The spin nutation resonance has been observed at a higher frequency than ferromagnetic resonance, e.g., while the ferromagnetic resonance occurs {in the} GHz regime, the nutation resonance occurs {in the} THz regime~\cite{neeraj2019experimental,Mondal2020nutation}. Moreover, the spin nutation shifts the ferromagnetic resonance frequency to a lower value.
{Although this shift} is very small, the line-width of the resonance decreases, however, and thus the effective damping {decreases,} too.

{S}pin nutation effects have not {yet comprehensively} %
been discussed in two-antiparallel aligned sublattice magnetic systems (e.g., antiferromagnets, ferrimagnets). In a recent investigation, it has been predicted that the spin nutation in antiferromagnets may have much significance~\cite{Mondal2020nutation}. Due to sublattice exchange interaction, the antiferromagnetic resonance frequency lies in the THz regime, while the nutation resonance frequency has similar order of magnitude. This helps to detect the antiferromagnetic {precession} and nutation resonances experimentally as they fall {in} the same frequency range. 
Moreover, the calculated shift of the antiferromagnetic resonance frequency is stronger than {that of a} ferromagnet. Additionally, the nutation resonance peak is exchange enhanced~\cite{Mondal2020nutation}, which is beneficial {for detection} in experiments. However, the previous investigation only considers the {\it intra}-sublattice inertial dynamics, while the effect of {\it inter}-sublattice inertial dynamics is unknown.

In {a} previous work, the LLG equation of motion with inter-sublattice Gilbert damping has been explored {by Kamra \textit{et al.}}~\cite{Kamra2018}. It {was} found that the introduction of inter-sublattice Gilbert damping enhances the damping~\cite{Kamra2018,Liu2017PRM,Yuan2019}. In this study, we formulate the spin dynamical equations in a two-sublattice magnetic system with {both} {\it intra} and {\it inter}-sublattice inertial dynamics {as well as \textit{inter} and \textit{intra}-sublattice Gilbert damping}, {extending thus previous work}~\cite{Mondal2020nutation}. First, we derive the magnetic susceptibility with the {\it inter}-sublattice effects and compute the precession and nutation resonance frequencies. We find that the precession resonance frequency and the effective Gilbert damping decrease with the intra-sublattice nutation {coupling} in antiferromagnets, however, {they} increase with the inter-sublattice nutation. {Unlike antiferromagnets, {we find for ferrimagnets that} the change of precession resonance frequencies is {more pronounced} with {both} {intra} and {inter}-sublattice nutation {coupling constants} in {the} exchange-like mode, {but} {nearly negligible for the} ferromagnetic mode.} 

The article is organized as follows. First, in Sec.\ \ref{2}, we discuss the linear-response theory of spin dynamics to calculate the {magnetic} susceptibility with the intra and inter-sublattice nutation effects. In Sec.\ \ref{3}, the precession resonance frequencies have been calculated with analytical and numerical tools for antiferromagnets (Sec.\ \ref{3a}) and ferrimagnets (Sec.\ \ref{3b}).
{We summarize the obtained results in}
Sec.\ \ref{4}.

\section{Linear-response {susceptibility} in two-sublattice magnets}
\label{2}
For two-sublattice magnetic systems, namely $A$ and $B$ representing the two sublattices, the ILLG equations of motion read
\begin{align}
\label{sublatticeA}
    \dot{\bm{M}}_{A} & = -\gamma_A \left(\bm{M}_A \times \bm{H}_A\right) + \frac{\alpha_{AA}}{M_{A0}}\left(\bm{M}_A\times \dot{ \bm{M}}_{A}\right) \nonumber\\
    & + \frac{\alpha_{AB}}{M_{B0}}\left(\bm{M}_A\times \dot{ \bm{M}}_{B}\right) + \frac{\eta_{AA}}{M_{A0}}\left(\bm{M}_A\times \ddot{ \bm{M}}_{A}\right) \nonumber\\
    & + \frac{\eta_{AB}}{M_{B0}}\left(\bm{M}_A\times \ddot{ \bm{M}}_{B}\right),\\
    \dot{\bm{M}}_{B} & = -\gamma_B \left(\bm{M}_B \times \bm{H}_B\right) + \frac{\alpha_{BB}}{M_{B0}}\left(\bm{M}_B\times \dot{ \bm{M}}_{B}\right) \nonumber\\
    & + \frac{\alpha_{BA}}{M_{A0}}\left(\bm{M}_B\times \dot{ \bm{M}}_{A}\right) + \frac{\eta_{BB}}{M_{B0}}\left(\bm{M}_B\times \ddot{ \bm{M}}_{B}\right) \nonumber\\
    & + \frac{\eta_{BA}}{M_{A0}}\left(\bm{M}_B\times \ddot{ \bm{M}}_{A}\right).
    \label{sublatticeB}
\end{align}
In the above dynamical equations, the first terms relate to the spin precession, the second and third terms represent the {\it intra} and {\it inter}-sublattice Gilbert damping, and the last two terms classify the {\it intra} and {\it inter}-sublattice inertial dynamics. The {\it intra}-sublattice magnetization dynamics has been characterized with the Gilbert damping {constants} $\alpha_{AA}$, $\alpha_{BB}$ and inertial relaxation time $\eta_{AA}$ or $\eta_{BB}$, while the {\it inter}-sublattice dynamics { is characterized by}  
Gilbert damping $\alpha_{AB}$ or $\alpha_{BA}$ and inertial relaxation time $\eta_{AB}$ or $\eta_{BA}$. Note that the Gilbert damping parameters are dimensionless, however, inertial relaxation time has a dimension of time \cite{Ciornei2010thesis,Ciornei2011,Mondal2017Nutation}. The extended equations of motions in Eqs.\ (\ref{sublatticeA}) and (\ref{sublatticeB}) represent general magnetization dynamics for two-sublattice magnets (e.g., antiferromagnets, ferrimagnets, two-sublattice ferromagnets, and so on). 

The free energy of the considered two-sublattice system reads 
\begin{align}
    \mathcal{F}\left(\bm{M}_{A}, \bm{M}_{B}\right)  & =  - H_0\left( M_{Az}+ M_{Bz}\right)  - \frac{K_A}{M^{2}_{A0}} M^2_{Az} \nonumber\\
    & - \frac{K_B}{M^{2}_{B0}} M^2_{Bz} + \frac{J}{M_{A0}M_{B0}}  \bm{M}_A\cdot \bm{M}_B\,. 
     \label{Free-energy}
\end{align}
Here, the first term defines the Zeeman coupling of two sublattice spins with an external field $\bm{H}_0 = H_0\hat{\bm{z}}$. The second and third terms represent the anisotropy energies for the sublattice $A$ and $B$, respectively. The last term can be identified as the Heisenberg exchange energy between the two sublattices. Note that the Heisenberg coupling energy, $J>0$ for antiferromagnets and ferrimagnets, however $J<0$ for ferromagnetic-like coupling.  

We calculate the effective field in the ILLG equation as the derivative of free energy in Eq.\ (\ref{Free-energy}) with respect to the corresponding magnetization
\begin{align}
\label{HeffA}
     \bm{H}_A  & =  -\frac{\partial \mathcal{F}\left(\bm{M}_A, \bm{M}_B\right)}{\partial \bm{M}_A}   \nonumber \\ 
     & = \left( H_0 + \frac{2K_A}{M^{2}_{A0}}M_{Az}\right)\hat{\bm{z}} 
      - \frac{J}{M_{A0}M_{B0}}\bm{M}_B\,,\\
     \bm{H}_B  & =  -\frac{\partial \mathcal{F}\left(\bm{M}_A, \bm{M}_B\right)}{\partial \bm{M}_B}  \nonumber \\
     & = \left( H_0 + \frac{2K_B}{M^{2}_{B0}}M_{Bz}\right)\hat{\bm{z}}
      - \frac{J}{M_{A0}M_{B0}}\bm{M}_A\,.
     \label{HeffB}
\end{align}
First, {in} the ground state, we consider that the $A$ sublattice magnetization is $\bm{M}_{A}  = M_{A0}\hat{\bm{z}}$, while the $B$ sublattice magnetization is antiparallel $\bm{M}_{B} = -M_{B0}\hat{\bm{z}}$, such that we can describe the antiferromagnets ($M_{A0} = M_{B0}$) and ferrimagnets ($M_{A0} > M_{B0}$).
We then expand the magnetization around the ground state in small deviations, $\bm{M}_{A} = M_{A0}\hat{\bm{z}}+\bm{m}_{A} (t)$ and $\bm{M}_{B} = -M_{B0}\hat{\bm{z}}+\bm{m}_{B}(t)$. The small deviations $\bm{m}_{A/B}$ are induced by the transverse external field $\bm{h}_{A/B}(t)$. 

{For convenience,} we work in the circularly polarized basis, {i.e.,}  $m_{A/B\pm} = m_{A/Bx} \pm \textrm{i} m_{A/By},\,\,h_{A/B\pm} = h_{A/Bx} \pm \textrm{i} h_{A/By}$, and define $\Omega_A = \gamma_A /M_{A0}( J + 2K_A + H_0 M_{A0}), \,\,\Omega_B = \gamma_B /M_{B0}
( J + 2K_B -  H_0 M_{B0})$. With the time-dependent {harmonic} fields and magnetizations $h_{A/B\pm},\,\,m_{A/B\pm} \propto e^{\pm {\rm i} \omega t}$, we obtain the {magnetic} susceptibility tensor \cite{Mondal2020nutation} 
\begin{widetext}
\begin{align}
    \begin{pmatrix}
     m_{A\pm}\\
     m_{B\pm}
    \end{pmatrix} & = \frac{1}{\Delta_\pm}\begin{pmatrix}
     \dfrac{1}{ \gamma_B M_{B0}}\left(\Omega_B \pm {\rm i}\omega\alpha_{BB} -\omega^2 \eta_{BB}+\omega  \right) &  - \dfrac{1}{\gamma_BM_{A0}} \left(\dfrac{ \gamma_B }{M_{B0}} J  \pm {\rm i} \omega \alpha_{BA}  - \omega^2 \eta_{BA} \right)\\
     - \dfrac{1}{\gamma_A M_{B0}} \left( \dfrac{ \gamma_A }{M_{A0}} J \pm{\rm i} \omega \alpha_{AB}  - \omega^2  \eta_{AB} \right) &   \dfrac{1}{ \gamma_A M_{A0}}\left(\Omega_A \pm {\rm i}\omega\alpha_{AA} -\omega^2 \eta_{AA} -\omega  \right)
    \end{pmatrix}\begin{pmatrix}
      h_{A\pm}\\
      h_{B\pm}
    \end{pmatrix}\nonumber\\
    & = \chi_{\pm}^{AB}\begin{pmatrix}
      h_{A\pm}\\
      h_{B\pm}
    \end{pmatrix}\,,
    \label{susceptibility}
\end{align}
with the definition of the determinant $\Delta_\pm = \left(\gamma_A  \gamma_B M_{A0}M_{B0}\right)^{-1}\left(\Omega_A \pm {\rm i}\omega\alpha_{AA} -\omega^2\eta_{AA}-\omega  \right)\left(\Omega_B \pm {\rm i}\omega\alpha_{BB} -\omega^2\eta_{BB} +\omega  \right) -   \left(\gamma_A  \gamma_B M_{A0}M_{B0}\right)^{-1}\left( \frac{\gamma_A}{M_{A0}}J \pm{\rm i} \omega \alpha_{AB}  - \omega^2  \eta_{AB} \right) \left(\frac{\gamma_B}{M_{B0}} J  \pm {\rm i} \omega \alpha_{BA}  - \omega^2 \eta_{BA} \right)$.
\end{widetext}
As one expects, the {\it inter}-sublattice Gilbert damping and inertial dynamical contributions arise in the off-diagonal components of the susceptibility tensor, while the {\it intra}-sublattice contributions are {in the} diagonal {component of the} susceptibility~\cite{Mondal2020nutation}. Note that without inertial dynamics terms, the expression {for the}  susceptibility {is} in accordance {with the one} derived by Kamra {\it et al.}~\cite{Kamra2018}. 

To find the resonance frequencies, the determinant $\Delta_\pm$ must go to zero, thus one has to solve the following fourth-order equation in frequency
\begin{equation}
\mathcal{A}_{\pm} \omega^4 + \mathcal{B}_{\pm} \omega^3 + \mathcal{C}_{\pm} \omega^2 + \mathcal{D}_{\pm} \omega + \mathcal{E}_{\pm}  = 0\,,
\label{4th-order}
\end{equation}
{where} the coefficients have the following forms
\begin{eqnarray}
\mathcal{A}_{\pm} & =&   \eta_{AA}\eta_{BB} - \eta_{AB}\eta_{BA}, \\
\mathcal{B}_{\pm} & =& \mp {\rm i} \left(\alpha_{AA}\eta_{BB}  + \alpha_{BB}\eta_{AA}\right) - \left(\eta_{AA} - \eta_{BB}\right)\nonumber\\
& & \pm {\rm i}\left(\alpha_{AB}\eta_{BA} + \alpha_{BA}\eta_{AB}\right), \\
\mathcal{C}_{\pm} & =& -1 \pm {\rm i} \left(\alpha_{AA} - \alpha_{BB}\right) - \left(\Omega_A\eta_{BB} + \Omega_B\eta_{AA}\right)\nonumber\\
& & - \alpha_{AA}\alpha_{BB} + \left(\frac{\gamma_A}{M_{A0}} \eta_{BA} + \frac{\gamma_B}{M_{B0}} \eta_{AB}\right) J \nonumber\\
& &+ \alpha_{AB}\alpha_{BA}, \\
\mathcal{D}_{\pm} & =& \left(\Omega_A - \Omega_B\right) \pm {\rm i} \left(\Omega_A\alpha_{BB} + \Omega_{B}\alpha_{AA}\right) \nonumber\\
& & \mp {\rm i} \left( \frac{\gamma_A}{M_{A0}} \alpha_{BA} + \frac{\gamma_B}{M_{B0}} \alpha_{AB}\right) J, \\
\mathcal{E}_{\pm} & =& \Omega_A\Omega_B - \frac{\gamma_A\gamma_{B}}{M_{A0}M_{B0}} J^2\,.
\end{eqnarray}
The solutions of the above equation (\ref{4th-order}) result in four different frequencies in {the} presence of a finite external field. Two of those frequencies can be associated with the magnetization precession resonance, $\omega_{\rm p\pm}$ (positive and negative modes) that exists even without nutation. The other two frequencies dictate the nutation resonance frequencies, $\omega_{\rm n\pm}$ (positive and negative modes). 

\section{Results and Discussion}
\label{3}
The intrinsic intra-sublattice inertial dynamics have been discussed extensively in Ref.\ \cite{Mondal2020nutation}. Essentially, the resonance frequencies and effective damping decrease with increasing intra-sublattice {inertial} relaxation time for antiferromagnets and ferrimagnets. Therefore, we consider a constant intra-sublattice {inertial} relaxation time in this work. 
In this section, we specifically discuss the effects of inter-sublattice nutation in {both} antiferromagnets and ferrimagnets.

\subsection{Antiferromagnets}
\label{3a}
\begin{figure}
    \centering
    \includegraphics[scale = 0.53]{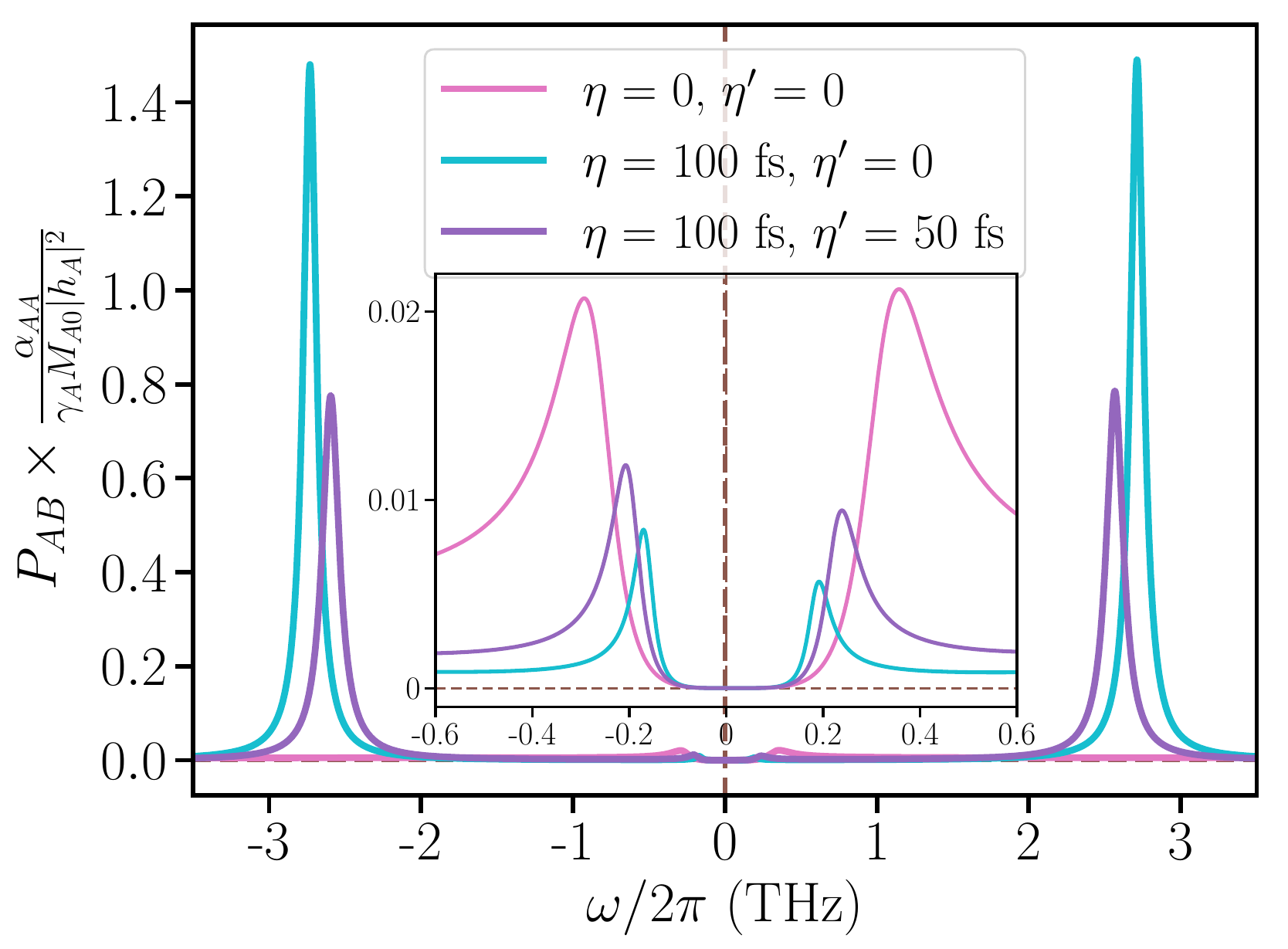}
    \caption{The calculated dissipated power vs.\ frequency for {an} antiferromagnet with $M_{A0} = M_{B0} = 2\, \mu_B$, {and various values of the intra- and inter-sublattice nutations parameters, $\eta$ and $\eta^{\prime}$}. The inset shows the dissipated power close to the precession resonance frequencies. The other used parameters are $\gamma_A = \gamma_B = 1.76\times 10^{11}$ T$^{-1}$s$^{-1}$, $J = 10^{-21}$ J, $K_A = K_B = 10^{-23}$ J, $H_0 = 1$ T, $\alpha_{AA} = \alpha_{BB} =  0.05$, $\alpha_{AB} = \alpha_{BA} = 0$, $\eta_{AA} = \eta_{BB} = \eta $ and $\eta_{AB} = \eta_{BA} = \eta^\prime$.}
    \label{Dissipated_power}
\end{figure}

\begin{figure*}[tbh!]
    \centering
    \includegraphics[scale = 0.27]{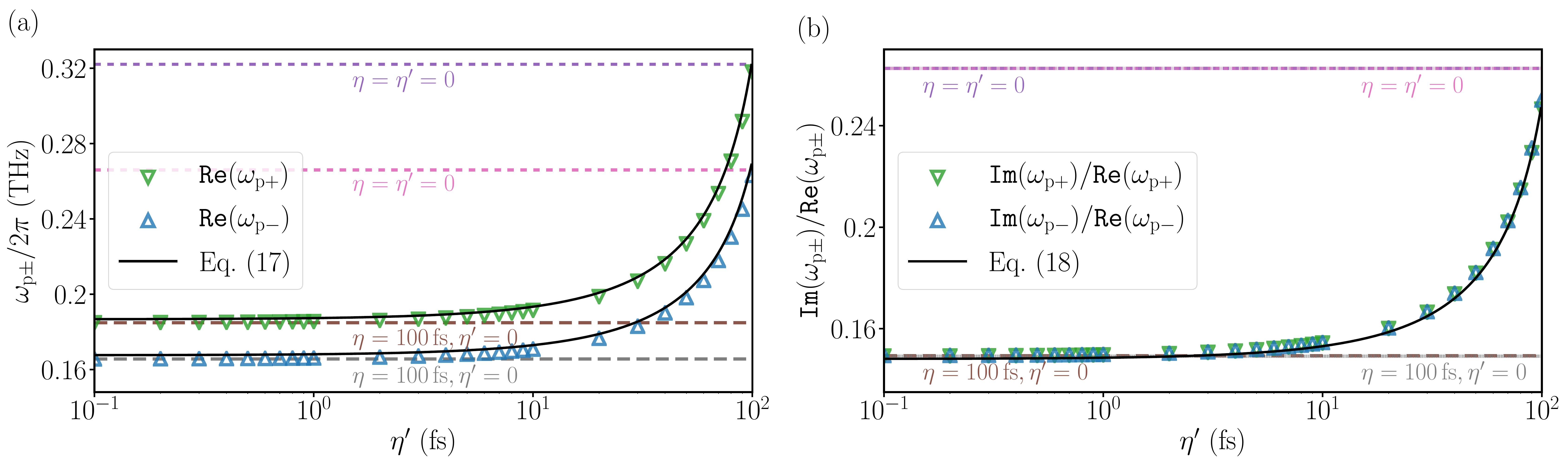}
    \caption{The calculated precession frequencies as a function of inter-sublattice nutation $\eta^{\prime}$ for {an} antiferromagnet, {setting} $M_{A0} = M_{B0} = 2\,\mu_B$. The data points denote the numerical solution of Eq.~(\ref{4th-order}) and the black lines correspond to the analytical solution in Eq.~(\ref{prece_freq}).  (a) The real part of the resonance frequency, and (b) the ratio of imaginary and real part of the frequency have been plotted. The other used parameters are $\gamma_A = \gamma_B = 1.76\times 10^{11}$ T$^{-1}$s$^{-1}$, $J = 10^{-21}$ J, $K_A = K_B = K = 10^{-23}$ J, $H_0 = 1$ T, $\alpha_{AA} = \alpha_{BB} = \alpha =  0.05$, $\alpha_{AB} = \alpha_{BA} = 0$, $\eta_{AA} = \eta_{BB} = \eta =  100$ fs and $\eta_{AB} = \eta_{BA} = \eta^\prime$. {The horizontal lines correspond to solutions with zero inter-sublattice nutation ($\eta' =0$).} {Note that we show $\tt{Re}(\omega_{\textrm{p}-})$ as $-\tt{Re}(\omega_{{\rm p}-})$.}}
    \label{second-order-freq-AFM}
\end{figure*}

{To start with, we calculate the frequency-dependent dissipated power of an antiferromagnet.}
Using the expressions {for the}  susceptibility in Eq.\ (\ref{susceptibility}), we calculate the dissipated power {in} the inertial dynamics with the following definition $P_{AB} = \dot{\bm{m}}_A\cdot \bm{h}_A +  \dot{\bm{m}}_B\cdot \bm{h}_B = \frac{1}{2} \left(\dot{m}_{A+}h_{A-}+\dot{m}_{A-}h_{A+} +\dot{m}_{B+}h_{B-}+\dot{m}_{B-}h_{B+}\right)$ which {leads to} a complicated expression {(not given)}. 
{For convenience, we define $\alpha_{AA} = \alpha_{BB} = \alpha$, $\eta_{AA} = \eta_{BB} = \eta$. To focus on the inter-sublattice nutation $\eta_{AB} = \eta_{BA} = \eta^\prime$, we set the inter-sublattice Gilbert damping to zero, i.e., $\alpha_{AB} = \alpha_{BA} = 0$, and choose $M_{A0}= M_{B0} = 2\, \mu_B$.}
The exchange and anisotropy energies, magnetic moments used in the {here-presented}  computations are {comparable} to a {typical} antiferromagnetic NiO~\cite{Hutchings1972,Baierl2016,Mondal2019PRB} or CoO~\cite{Archer2008,Archer2011} system. However, we mention that  NiO or CoO {bulk crystals have} biaxial anisotropy. Also, the Gilbert damping {of} NiO is very small $\alpha \sim 10^{-4}$, {i.e., less than the here-used value.} In contrast, a large spin-orbit coupling in antiferromagnetic CrPt {(that has $\sim$2\,$\mu_B$ Cr moments) leads to a} higher Gilbert damping $\alpha \sim 10^{-2}$~\cite{Besnus1973,Zhang2012JAP}. {Importantly,} the inertial relaxation times $\eta$ and $\eta^\prime$ are not known in these antiferromagnetic systems. {Our simulations pertain therefore to typical, selected model systems.}
We show the evaluated dissipated power with and without inertial dynamics for {such} antiferromagnet in Fig.\ \ref{Dissipated_power}. Note that the dissipated power has already been calculated in Ref.~\cite{Mondal2020nutation}, however, without the {\it inter}-sublattice inertial dynamics.
{We can} observe that while the {\it intra}-sublattice inertial dynamics decreases the precessional resonance frequencies (see the {cyan lines} in Fig.\ \ref{Dissipated_power}), the {\it inter}-sublattice inertial dynamics works oppositely. Note that the nutation resonance frequencies decrease with the introduction of {\it inter}-sublattice inertial dynamics.

{To} understand the effect of {the} inter-sublattice nutation terms, first, we solve 
the Eq.~(\ref{4th-order}), considering {again} $\alpha_{AA} = \alpha_{BB} = \alpha$, $\eta_{AA} = \eta_{BB} = \eta$,  $\eta_{AB} = \eta_{BA} = \eta^\prime$, and $\alpha_{AB} = \alpha_{BA} = 0$. As the nutation in antiferromagnets is exchange enhanced~\cite{Mondal2020nutation}, we calculate the effect of inter-sublattice terms on {the} precession and nutation frequencies, {setting}
 $\gamma_A = \gamma_B = \gamma$ and $M_{A0} = M_{B0} = M_0$ {for antiferromagnets}. Therefore, the fourth-order equation in Eq.\ (\ref{4th-order}) reduces to an equation with $\mathcal{A}_{\pm}^{\rm AFM} = \eta^2 -\eta^{\prime 2} $, $\mathcal{B}_\pm^{\rm AFM} = \mp {\rm i} 2\alpha \eta $, $\mathcal{C}_{\pm}^{\rm AFM} = -1  - \left(\Omega_A + \Omega_B\right)\eta + 2\frac{\gamma}{M_{0}} \eta^\prime J 
 $, $\mathcal{D}_{\pm}^{\rm AFM} = \left(\Omega_A - \Omega_B\right) \pm {\rm i} \left(\Omega_A + \Omega_{B}\right)\alpha$, and $\mathcal{E}_\pm^{\rm AFM} = \Omega_A\Omega_B - \left(\frac{\gamma}{M_{0}}J\right)^2$.
  The solution of the above equation results in precession and nutation resonance frequencies for the two modes (positive and negative). Inserting the real and imaginary parts of the solutions $\omega_\pm = {\tt Re}\left(\omega_\pm\right) + {\rm i} {\tt Im}(\omega_\pm)$, we numerically calculate the precession resonance frequencies and effective damping (the ratio of imaginary and real frequencies) for {an} antiferromagnet as a function of inter-sublattice nutation.
  {The results are shown} in Fig.\ \ref{second-order-freq-AFM}, {where} the data points correspond to the numerical solutions.

 On the other hand, the fourth-order equation, $\mathcal{A}_+^{\rm AFM} \omega^4+\mathcal{B}_+^{\rm AFM} \omega^3+\mathcal{C}_+^{\rm AFM} \omega^2 +\mathcal{D}_+^{\rm AFM} \omega + \mathcal{E}_+^{\rm AFM} = 0$ can analytically be solved using the considerations that $K_A = K_B = K$, $J\gg K$, $M_0H_0$ and $\alpha \ll 1$. Therefore, one has $\Omega_A = \Omega_B \approx \gamma (J+2K)/M_0$. Essentially, the fourth-order equation reduces to 
 \begin{align}
    & \left(\eta^2 -\eta^{\prime 2}\right)\omega^4 -\left[1 +2\frac{\gamma \eta (J+2K)}{M_0} -2\frac{\gamma \eta^\prime J}{M_0}  \right] \omega ^2\nonumber\\
    & - 2{\rm i} \alpha \eta \omega_{(0)}^3 + 2\gamma H_0 \omega_{(0)}+  \frac{2{\rm i}\gamma \alpha}{M_0}\left(J+2K\right) \omega_{(0)}\nonumber\\
     &  +\frac{\gamma^2}{M_0^2} \left(J+2K\right)^2 - \frac{\gamma^2 J^2}{M_0^2}-  \gamma^2 H_0 ^2  = 0\,,
     \label{2nd_order_omega0}
 \end{align}
 {with} $\omega_{(0)}$ being the solution of the above equation {for} $\alpha = 0$ and $H_0 = 0$.
 The solutions of the above equation are rather simple {and} provide the two precession frequency modes (positive and negative) for antiferromagnets. Expanding the solutions of Eq.\ (\ref{2nd_order_omega0}) up to the first order in $\alpha$ and $H_0$, and also in first order in $K/J \ll 1$, the precession resonance frequencies are obtained (neglecting the higher-order in $\omega_{(0)}$-{terms}) {as}
 \begin{align}
     \omega_{\rm p\pm} & \approx \pm \frac{\gamma}{M_0} \frac{\sqrt{4K\left(K+J\right)}}{\sqrt{1+ \dfrac{4\gamma \eta K}{M_0} + \dfrac{2\gamma J}{M_0} \left(\eta - \eta^\prime \right)}}\nonumber\\
     & + \frac{\gamma H_0 +  {\rm i}  \dfrac{\gamma\alpha}{M_0}\left(J+2K\right) }{\sqrt{1+ \dfrac{4\gamma \eta K}{M_0} + \dfrac{2\gamma J}{M_0} \left(\eta - \eta^\prime \right)}} ~\frac{\vert \omega_{(0)}\vert }{\dfrac{\gamma}{M_0}\sqrt{4K(K+J)}}.
     \label{prece_freq}
 \end{align}
Now substituting the $\vert \omega_{(0)} \vert $ from the leading term in the frequency expression into the perturbative terms in Eq.\ (\ref{prece_freq}), the approximate precession frequencies are {obtained as}
\begin{eqnarray}
    \omega_{\rm p\pm} & \approx & \pm \frac{\gamma}{M_0} \frac{\sqrt{4K\left(K+J\right)}}{\sqrt{1+ \dfrac{4\gamma \eta K}{M_0} + \dfrac{2\gamma J}{M_0} \left(\eta - \eta^\prime \right)}}\nonumber\\
     & & + \frac{\gamma H_0+{\rm i} \dfrac{\gamma\alpha}{M_0}\left(J+2K\right)}{1+ \dfrac{4\gamma \eta K}{M_0} + \dfrac{2\gamma J}{M_0} \left(\eta - \eta^\prime \right)} .
     \label{prec_freq_final}
\end{eqnarray}
This equation has been plotted in Fig.\ \ref{second-order-freq-AFM} as black lines.
{Note that, for $\omega_{\rm p-}$ we show for convenience $-\tt{Re}(\omega_{\rm p-})$ in Fig.\  \ref{second-order-freq-AFM}(a) and in the following.}
Due to the presence of $\eta - \eta^\prime$ in the denominator of the frequency expressions, the precession resonance frequency increases when inter-sublattice nutation is taken into account ($\eta^\prime < \eta$), which explains the increase in frequency in Fig.\ \ref{second-order-freq-AFM}(a). At the limit $\eta \rightarrow \eta^\prime $, the nutation (intra and inter-sublattice) does not play a significant role as {the} precession resonance frequency is decreased by a factor $\sqrt{1+\frac{4\gamma \eta K}{M_0}}$ which is very small due to $K\ll J$. Note that the two resonance frequencies are approximately 0.332 THz and 0.276 THz with $\alpha = 0$ and $\eta = \eta^\prime = 0$, while these two frequencies are 0.322 THz and 0.266 THz with $\alpha = 0.05$ and  $\eta = \eta^\prime = 0$. The latter has been shown in Fig.\ \ref{second-order-freq-AFM}(a) as dashed lines. Therefore, the Gilbert damping has already {the} effect {that it} reduces the resonance frequencies. 

The effective Gilbert damping can be calculated using the ratio between the imaginary and real parts of the frequencies, {i.e., the line width}. From Eq.\ (\ref{prec_freq_final}) one arrives {at}
\begin{align}
    \frac{{\tt Im}\left(\omega_{\rm p\pm}\right)}{{\tt Re}\left(\omega_{\rm p \pm}\right)} \! & \approx \alpha \frac{(J+2K)}{\sqrt{4K(K+J)}} \frac{1}{\sqrt{1+ \dfrac{4\gamma \eta K}{M_0} + \dfrac{2\gamma J}{M_0} \left(\eta - \eta^\prime \right)}}\,.
    \label{AFM_Damping}
\end{align}
{Note that the two resonance modes have the same effective damping. For ferromagnets, the exchange energies do not contribute and thus the effective damping remains the same as $\alpha$, in the absence of magnetic inertial terms (see \cite{MondalJPCM2021}). However, in antiferromagnets the effective damping is enhanced due to the exchange interaction by a factor $\frac{(J+2K)}{\sqrt{4K(K+J)}}$, even without any inertial terms. As investigated earlier \cite{Mondal2020nutation}, the effective damping decreases with the intra-sublattice relaxation time. However,}
similar to the increase in frequency, the effective damping also increases with the inter-sublattice inertial relaxation time, as seen in Fig.\ \ref{second-order-freq-AFM}(b). {The analytical solution in Eq.\ (\ref{AFM_Damping}) agrees excellently with the numerical solutions.}  Close to the limit $\eta^\prime \rightarrow \eta$, the effective Gilbert damping in Eq.\ (\ref{AFM_Damping})  one expects the effective damping to be {in}creased by a factor $\left({1+\frac{4\gamma \eta K}{M_0}}\right)^{-1/2}$, 
{as can be seen in Fig.\ \ref{second-order-freq-AFM}(b).}
 
\begin{figure}[tbh!]
    \centering
    \includegraphics[scale = 0.3]{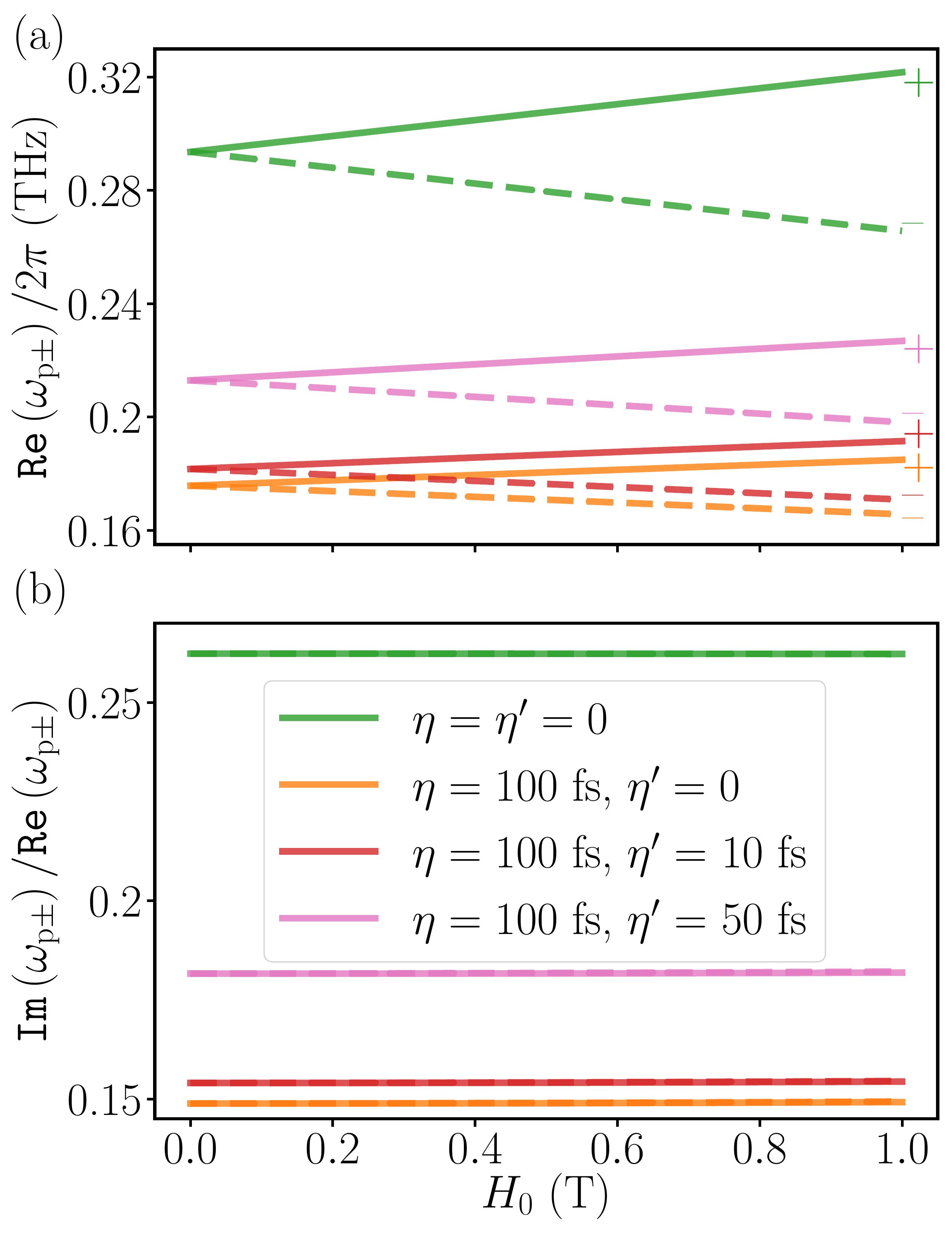}
    \caption{The calculated precession frequencies at several inter-sublattice relaxation times as a function of applied field for antiferromagnets using $M_{A0} = M_{B0} = 2\,\mu_B$. The solid and dashed lines represent the positive and negative modes, respectively. (a) The real part of the resonance frequencies and (b) the ratio of imaginary and real part of the frequency have been plotted. The other used parameters are $\gamma_A = \gamma_B = 1.76\times 10^{11}$ T$^{-1}$s$^{-1}$, $J = 10^{-21}$ J, $K_A = K_B = K = 10^{-23}$ J, $\alpha_{AA} = \alpha_{BB} = \alpha = 0.05$, $\alpha_{AB} = \alpha_{BA} = 0$, $\eta_{AA} = \eta_{BB} = \eta =  100$ fs and $\eta_{AB} = \eta_{BA} = \eta^\prime$.
    }
    \label{fourth-order-freq-AFM}
\end{figure}
\begin{figure}[tbh!]
    \centering
    \includegraphics[scale = 0.3]{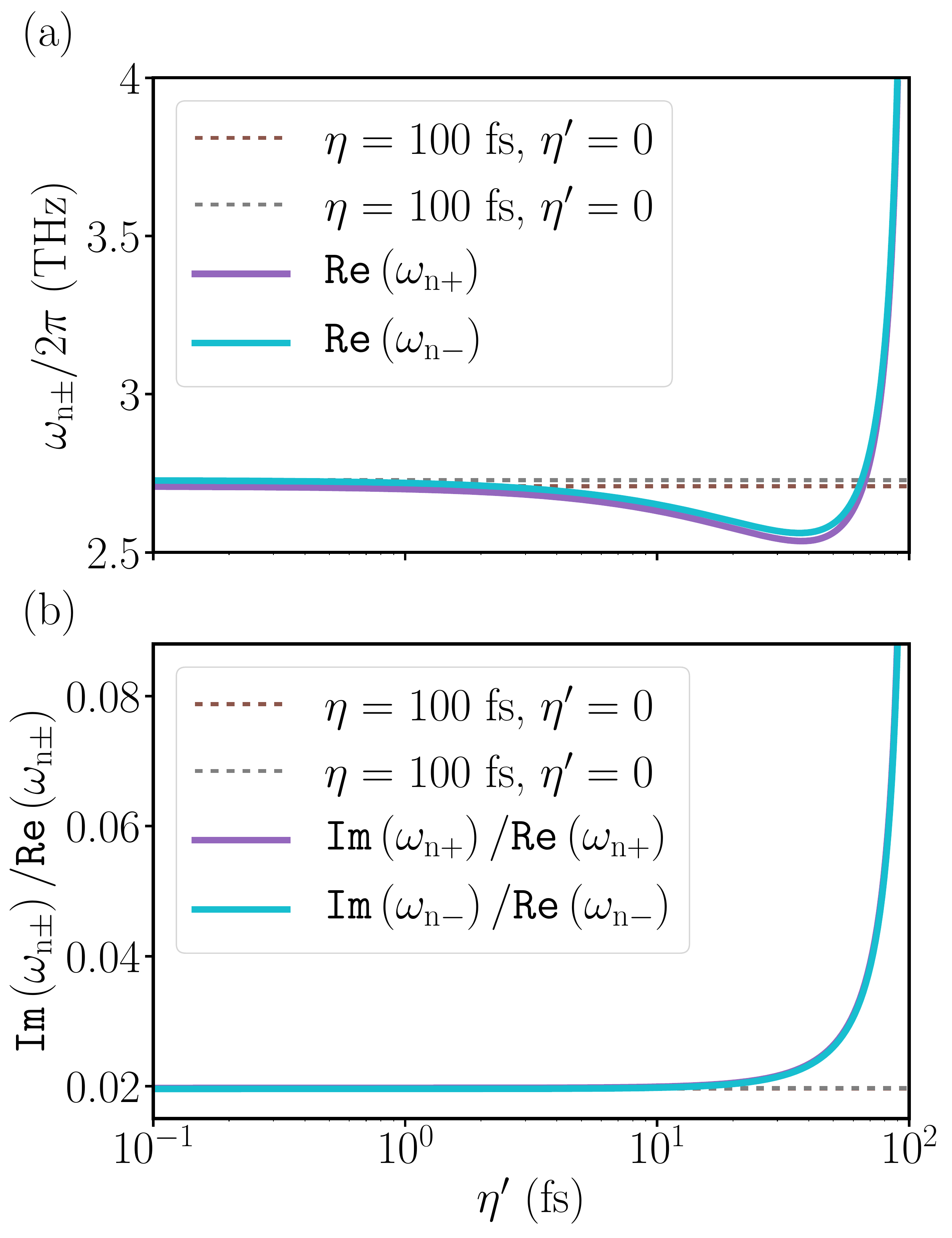}
    \caption{The calculated nutation frequencies as a function of inter-sublattice nutation for antiferromagnets using $M_{A0} = M_{B0} = 2\,\mu_B$. (a) The real part of the nutation resonance frequencies and (b) the ratio of imaginary and real part of the nutation resonance frequency have been plotted. The other used parameters are $\gamma_A = \gamma_B = 1.76\times 10^{11}$ T$^{-1}$s$^{-1}$, $J = 10^{-21}$ J, $K_A = K_B = K = 10^{-23}$ J, $H_0 = 1$ T, $\alpha_{AA} = \alpha_{BB} = \alpha = 0.05$, $\alpha_{AB} = \alpha_{BA} = 0$, $\eta_{AA} = \eta_{BB} = \eta =  100$ fs and $\eta_{AB} = \eta_{BA} = \eta^\prime$.}
    \label{fourth-order-nut-freq-AFM}
\end{figure}

Next, we discuss the field dependence of {the} resonance frequencies. The precession resonance frequencies and effective damping have been plotted as a function of the applied field $H_0$ {for} several inter-sublattice relaxation times in Fig.\ \ref{fourth-order-freq-AFM}. As {can be} observed, at zero applied field, the two modes (positive and negative) coincide in antiferromagnets, a fact that can be seen from Eq.\ (\ref{prec_freq_final}). 
However, the applied field {induces} the splitting of these two modes. The frequency splitting scales with $\left[1+ \dfrac{4\gamma \eta K}{M_0} + \dfrac{2\gamma J}{M_0} \left(\eta - \eta^\prime \right)\right]^{-1} \! \gamma H_0$, meaning that the splitting is linear in {the}  applied field, $H_0$. On the other hand, at a constant field, the splitting also depends on the inter- and intra-sublattice nutation. From Eq.\ (\ref{prec_freq_final}), it is clear that the splitting is reduced with intra-sublattice nutation, while it is enhanced with inter-sublattice nutation. Such a conclusion can also be {drawn from the} numerical solutions in Fig.\ \ref{fourth-order-freq-AFM}(a). The effective damping of the antiferromagnet remains field independent which can be observed in Fig.\ \ref{fourth-order-freq-AFM}(b).

{Proceeding as} previously, we obtain the following nutation frequencies
\begin{align}
    \omega_{\rm n\pm} & \approx \pm \frac{1}{\eta} \sqrt{\dfrac{1 +\dfrac{4 \gamma \eta K}{M_0} + \dfrac{2\gamma J}{M_0}\left(\eta -  \eta^\prime\right)}{1 - \dfrac{\eta^{\prime 2}}{\eta^2}}}\Bigg(1 -\nonumber\\
    &  \dfrac{\dfrac{(\eta^2 -\eta^{\prime 2})\gamma ^2}{M_0} \times 4K(J+K)}{2\left[1 +\dfrac{4\gamma\eta  K}{M_0} + \dfrac{2\gamma  J}{M_0} \left(\eta - \eta^\prime\right)\right]^2} \Bigg)\nonumber\\
    & - \dfrac{\gamma H_0 - {\rm i}\alpha\left[\dfrac{\eta}{\eta^2 - \eta^{\prime 2}} + \dfrac{\gamma}{M_0} \left(J+2K\right)\right] }{1 +\dfrac{4\gamma \eta K}{M_0} + \dfrac{2\gamma  J}{M_0} \left(\eta - \eta^\prime\right)}\,.
    \label{Nut_freq}
\end{align}

Note that at the limit $ \eta^{\prime} \rightarrow 0$, the nutation frequencies without the inter-sublattice coupling are {recovered}
\cite{Mondal2020nutation}. The dominant term in the calculated frequency is the first term in Eq.\ (\ref{Nut_freq}). With the introduction of inter-sublattice coupling $\eta^{\prime}$, both the numerator and denominator of the dominant frequency term decrease 
{and therefore,} 
the nutation frequencies approximately stay constant (with a slow decrease) with inter-sublattice nutation when $\eta> \eta^\prime$ as plotted in Fig.\ \ref{fourth-order-nut-freq-AFM}. However, {in} the limit $\eta^\prime \rightarrow \eta$, the denominator vanishes, and thus the nutation frequencies diverge {as} can be seen in Fig.\ \ref{fourth-order-nut-freq-AFM}. It is interesting to note that the inter-sublattice inertial dynamics increase the precession resonance frequencies, however, decrease the nutation frequency. Such observation is also consistent with the dissipated power in Fig.~\ref{Dissipated_power}. 
The damping of the inertial dynamics also shows {a} similar behavior: {it} stays {nearly} constant with {a} divergence at the limit $\eta^\prime \rightarrow \eta$.

{As mentioned before,} the inertial relaxation times $\eta$ and $\eta^\prime$ are not known in for typical  antiferromagnetic systems.  {Notwithstanding}, we {obtain the general result} that the precession resonance frequencies decrease with {\it intra}-sublattice inertial dynamics, however, increase with {\it inter}-sublattice inertial dynamics. Thus, to experimentally realize the signature of inertial dynamics, an antiferromagnet with a higher ratio of intra to inter-{sublattice} inertial relaxation time ($\eta/\eta^\prime \gg 1$) is better suited.    

\subsection{Ferrimagnets}
\label{3b}
\begin{figure}
    \centering
    \includegraphics[scale = 0.32]{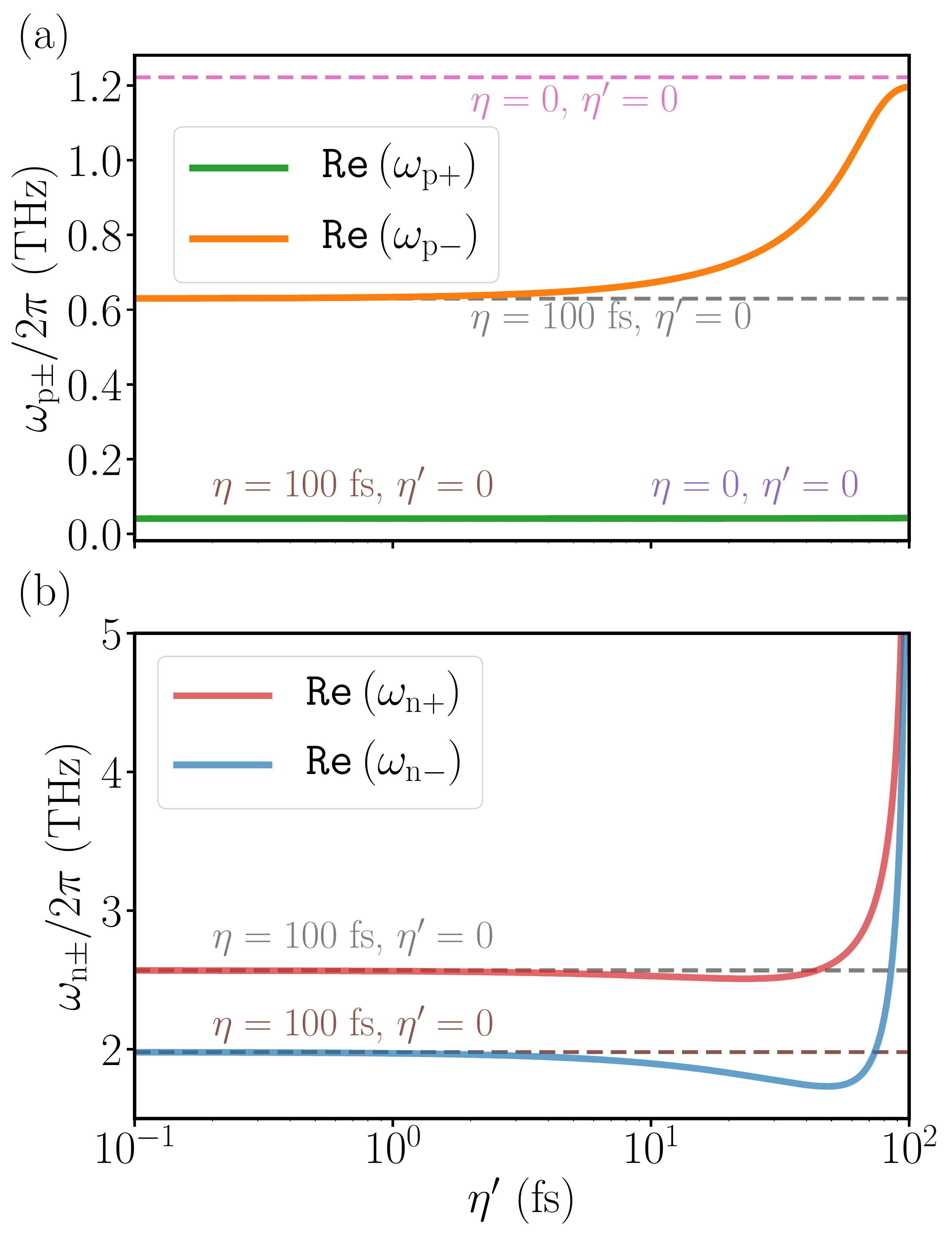}
    \caption{The calculated precession and nutation frequencies as a function of inter-sublattice nutation for ferrimagnets using $M_{A0} = 5 M_{B0} = 10\,\mu_B$.  The real part of the  (a) precession resonance  and (b) nutation resonance frequencies have been plotted. The other used parameters are $\gamma_A = \gamma_B = 1.76\times 10^{11}$ T$^{-1}$s$^{-1}$, $J = 10^{-21}$ J, $K_A = K_B = K = 10^{-23}$ J, $H_0 = 1$ T, $\alpha_{AA} = \alpha_{BB} = \alpha =  0.05$, $\alpha_{AB} = \alpha_{BA} = 0$, $\eta_{AA} = \eta_{BB} = \eta =  100$ fs and $\eta_{AB} = \eta_{BA} = \eta^\prime $.}
    \label{ferrimagnet_freq}
\end{figure}
Next, we consider {a} ferrimagnetic system where the magnetic moments in {the} two sublattices are different, i.e., $M_{A0}\neq M_{B0}$. In this case, the analytical solution of Eq.~(\ref{4th-order}) becomes cumbersome. The main reason is that $\Omega_A \neq \Omega_B$ for ferrimagnets, in fact, we calculate $\Omega_A - \Omega_B = \frac{\gamma \left(J + 2K\right)\left(M_{A0} - M_{B0}\right)}{M_{A0}M_{B0}} + 2\gamma H_0 $. For antiferromagnets, the magnetic moments in {the} two sublattices are exactly the same, i.e., $M_{A0} = M_{B0}$ and thus, within the approximation of $J\gg M_0H_0$, we find $\Omega_A = \Omega_B$ which simplifies the analytical solution of Eq.~(\ref{4th-order}). Thus, we numerically solve the Eq.~(\ref{4th-order}) to calculate the precession and nutation resonance frequencies for ferrimagnets. 
{We consider the case where $M_{A0} = 10\,\mu_B$ and $M_{B0}= 2\,\mu_B$, reminiscent of rare-earth--transition-metal ferrimagnets as GdFeCo \cite{Stanciu2007,Vahaplar2009} or TbCo \cite{Alebrand2012,Mangin2014,Ciuciulkaite2020}. However, we emphasize that the inertial relaxation times $\eta$ and $\eta^\prime$ are not known for these materials.}
The calculated precession frequencies {are} shown in Fig.~\ref{ferrimagnet_freq}{\blue (a)}. The effect of intra-sublattice inertial dynamics has already been studied in Ref.~\cite{Mondal2020nutation}. For ferrimagnets, the negative frequency mode appears to have a higher frequency {(i.e., larger negative)} than the positive one. However, both precession frequencies decrease with intra-sublattice relaxation time, $\eta$ \cite{Mondal2020nutation}. We, therefore, have set the intra-sublattice relaxation time $\eta$ to 100 fs and vary the inter-sublattice relaxation time $\eta^\prime<\eta$.  
The upper precession resonance mode $\omega_{\rm p-}$ -- {the} exchange-like mode -- increases with the inter-sublattice relaxation time $\eta^{\prime}$, while the ferromagnetic-like mode $\omega_{\rm p+}$ shows {a} very small increase. {Thus,} for ferrimagnets, the change in precession frequencies is {more significant} in the exchange-like mode than {in} the ferromagnetic-like mode. At the limit $\eta^\prime \rightarrow \eta$, the {precession} resonance frequencies almost coincide with the resonance frequencies calculated at $\eta = \eta^\prime = 0$, meaning that the inertial dynamics do not play any role {for} the {precession} resonance frequency. The latter can clearly be seen in Fig.\ \ref{ferrimagnet_freq}(a). These observations are similar to the antiferromagnet as discussed earlier.  The nutation resonance frequencies in Fig.\ \ref{ferrimagnet_freq}(b) again decline with the inter-sublattice relaxation time showing a divergence at the limit $\eta^\prime\rightarrow\eta$. 
{However, one can notice {here} two distinguishable nutation resonance frequencies unlike {almost a single-valued nutation frequencies of} antiferromagnets.}

\section{Summary}
\label{4}
In summary, we have formulated a linear-response theory of {the} ILLG equations for antiferromagnets with inter- and intra-sublattice inertial dynamics. The calculation of {the} susceptibility tensor shows that the intra-sublattice terms appear in the diagonal elements, while the inter-sublattice terms appear in the off-diagonal elements. The dissipated power contains a precession resonance peak {in} the {sub-}THz regime for antiferromagnets, however, the introduction of inertial dynamics {causes} another peak, a nutation resonance peak at a higher, {few THz} frequency. Moreover, we observe that the inter-sublattice inertial dynamics work oppositely to the {intra}-sublattice inertial one. By finding the poles of the susceptibility, we calculate the precession and nutation resonance frequencies. While the {precession} resonance frequencies decrease with intra-sublattice relaxation time, the inter-sublattice inertial dynamics have the opposite effect. In fact, we observe that the magnetic inertia does not have any effect on {the} antiferromagnetic {precession} resonance at the limit $\eta^\prime \rightarrow\eta$. On the other hand, {the THz} nutation resonance frequency decreases {slightly} with the introduction of inter-sublattice inertial dynamics, however, showing a divergence at the limit $\eta^\prime \rightarrow\eta$. Our derived analytical theory explains such inter-sublattice contributions. Finally, for ferrimagnets, we find a similar behavior for the inter-sublattice inertial dynamics. {However, the precession resonance frequency of the exchange-like mode depends significantly on the nutation couplings in contrast to that of the ferromagnetic-like mode that is practically independent of the nutation constants.}

\section*{Acknowledgments}
We acknowledge Levente R\'ozsa and Ulrich Nowak for fruitful discussions, the Swedish Research Council (VR Grant No.\ 2019-06313) for research funding {and Swedish National Infrastructure for Computing (SNIC) at
NSC Link{\"o}ping for computational resources}. {We further acknowledge support through the European Union's Horizon2020 Research and Innovation Programme under Grant agreement No.\ 863155 (s-Nebula).}

\bibliographystyle{apsrev4-1}
%\bibliography{Ref}

%\end{document}


\end{document}